\documentclass{acmsiggraph}
\usepackage{amsfonts,empheq}
\usepackage{caption}
\usepackage{subcaption}
\usepackage{gensymb}
\usepackage{paralist}
\usepackage{booktabs}
\usepackage{graphicx}
\usepackage{float}
\usepackage{tikz}
\usetikzlibrary{3d}
\usetikzlibrary{external}
\usepackage{algorithm}
\usepackage[noend]{algpseudocode}
\usepackage{gensymb}
\newcommand\numberthis{\addtocounter{equation}{1}\tag{\theequation}}
\newcommand*\diff{\mathop{}\!\mathrm{d}}

\title{Sampling BSSRDFs with non-perpendicular incidence}

\author{
	Etienne Ferrier\thanks{e-mail: etiferrier@gmail.com}\\EPFL
}
\pdfauthor{Etienne Ferrier}

\TOGonlineid{45678}

\begin{document}

 \teaser{
 	\begin{subfigure}[b]{0.31\textwidth}
        \includegraphics[width=\textwidth]{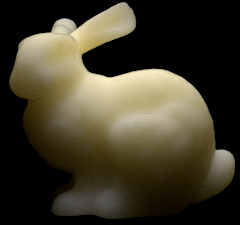}
        \caption{BSSRDF model assuming perpendicular incidence}
    \end{subfigure}%
    \begin{subfigure}[b]{0.31\textwidth}
        \includegraphics[width=\textwidth]{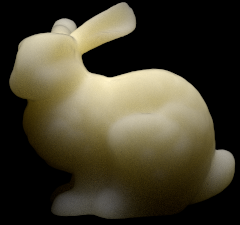}
        \caption{Our BSSRDF model}
    \end{subfigure}%
    \begin{subfigure}[b]{0.31\textwidth}
    	\includegraphics[width=\textwidth]{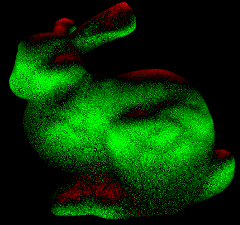}
        \caption{Luminance variations}
    \end{subfigure}
    \begin{subfigure}[b]{0.045\textwidth}
    	\includegraphics[width=\textwidth]{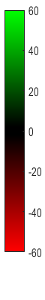}
    \end{subfigure}

   \caption{Comparison between our multiple-scattering BSSRDF model which accounts for non-perpendicular incidence (b) and a BSSRDF model assuming perpendicular incidence (a).  Luminance (R+G+B) variations between the two images are shown in (c). Our model scatters light further down on the side of the bunny and makes the top of the ears more transparent. The scene is the Stanford Bunny lit with a diffuse area light source from above. The material used is white grapefruit juice from \protect\cite{narasimhan}. The images were rendered using particle tracing in PBRT.}
   \label{render_bunny}
}
\maketitle

\begin{abstract}

Sub-surface scattering is key to our perception of translucent materials. Models based on diffusion theory are used to render such materials in a realistic manner by evaluating an approximation of the  material BSSRDF at any two points of the surface. Under the assumption of perpendicular incidence, this BSSRDF approximation can be tabulated over 2 dimensions to provide fast evaluation and importance sampling. However, accounting for non-perpendicular incidence with the same approach would require to tabulate over 4 dimensions, making the model too large for practical applications.
In this report, we present a method to efficiently evaluate and importance sample the multi-scattering component of diffusion based BSSRDFs for non-perpendicular incidence. Our approach is based on tabulating a compressed angular model of Photon Beam Diffusion. We explain how to generate, evaluate and sample our model. We show that 1 MiB is enough to store a model of the multi-scattering BSSRDF that is within $0.5\%$ relative error of Photon Beam Diffusion. Finally, we present a method to use our model in a Monte Carlo particle tracer and show results of our implementation in PBRT.

\end{abstract}


\keywordlist

\conceptlist

\section{Introduction}

\begin{figure}
\centering
\begin{subfigure}{.4\linewidth}
  	\centering
	\includegraphics[width=.6\textwidth]{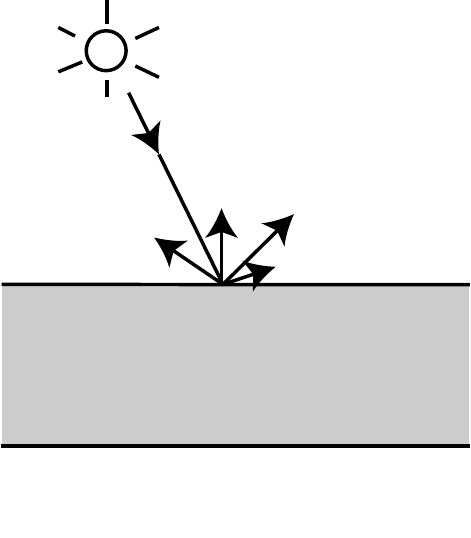}
  	\caption{Opaque material}
\end{subfigure}
\begin{subfigure}{.4\linewidth}
 	\centering
	\includegraphics[width=.6\textwidth]{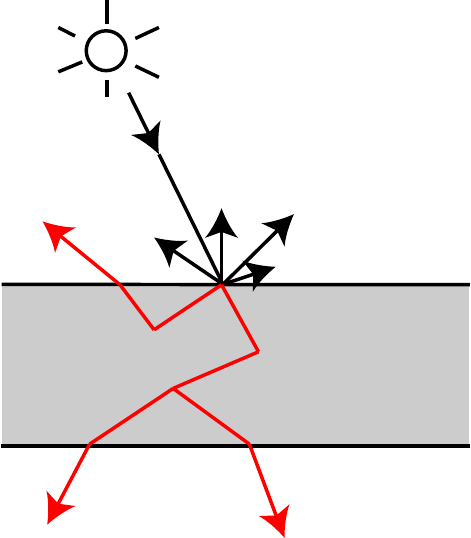}
 	\caption{Translucent material}
\end{subfigure}

\caption{Differences in light patterns between an opaque material and a translucent material. The gray area represents the material. In the case of an opaque material (a), light is either absorbed or reflected. Whereas in the case of a translucent material (b), a fraction of the incoming light is scattered below the surface where it continues to be scattered until exiting the material. The red light path in (b) shows the additional pattern responsible for translucency.}
\label{fig:fig_trans_schematics}
\end{figure}

Materials such as wax, marble and skin scatter light below their surface. This phenomenon plays an important role in their appearance, which is why physics based renderers aim to reproduce it. Figure~\ref{fig:fig_trans_schematics} illustrates this behavior by comparison with opaque materials.

In Monte-Carlo based path tracers, there exist several ways to render sub-surface scattering realistically. Volumetric path tracing, for instance, extends path tracing to the interior of objects to simulate internal scattering. However, this technique can be slow, especially for materials with a high amount of scattering. Other techniques, based on diffusion theory, aim to approximate the BSSRDF of the material under the assumption of a large number of internal bounces (multi-scattering). Several settings exist to solve the diffusion equation, such as using finite elements method on a discretization of the interior of objects. In this project, we focus on dipole based methods which provide an analytic approximation of the BSSRDF in a planar setting. But while these methods provide a way to evaluate the BSSRDF, they do not provide a specific scheme to importance sample it.

In PBRT, a tabulated model of the BSSRDF is generated for every sub-surface scattering material in the scene during a pre-processing phase. This model allows fast importance sampling and evaluation  of the BSSRDF using interpolation. However, tabulating every dimension of the BSSRDF would require too much memory. Thus in practice only the BSSRDF resulting from perpendicular incidence is tabulated, causing anisotropic effects -- which result from the diffusion of light arriving at oblique angles of incidence -- to be ignored.

In this report, we present a method to compute a compact model of non-perpendicular multi-scattering BSSRDFs that provides fast evaluation and importance sampling. Our approach expands PBRT tabulation approach with an analytic model of azimuthal variations. As in PBRT, we rely on Photon Beam Diffusion to generate our model during pre-processing. We show that our method provides a compact (1 Mib) and accurate ($0.5\%$ relative error) representation of Photon Beam Diffusion, even for grazing angles. We show images rendered using our BSSRDF model with particle tracing in PBRT and highlight differences with the previous implementation. Finally, we explain what remains to be done in order to use our model for path tracing.

\subsection{Related work}

Our approach is inspired by and expands on the BSSRDF implementation in PBRT~\cite{pbrt}. We also use PBRT as a framework to render images using our technique. In this document, we will only focus on the multi-scattering component of BSSRDFs. Treatment of the single-scattering component is left as future work.

Photon Beam Diffusion~\cite{habel2013photon} provides an analytic expression of multi-scattering BSSRDFs in a planar setting. We use this expression both as a reference for our model and a tool to generate it. However our goal is not only to build a compact model of Photon Beam Diffusion evaluation function but also to provide an efficient way to sample it.

\subsection{Contributions}

\begin{itemize}
\item We provide a method to compute a compact model of the multiscattering component of diffuse BSSRDFs that accounts for oblique incidence and allows fast importance sampling 
\item We explain how to use our model in a particle tracer
\end{itemize}

\section{Sub-surface scattering}

\begin{table}
\centering
\begin{tabular}{ c l c }
  \hline
  \vspace{-0.3cm}\\
  Symbol & Description & Units \\
  \hline
  $g$ & Average cosine of scattering & - \\
  $\eta$ & Relative index of refraction & - \\
  $\sigma_s$ & Scattering coefficient & [m\textsuperscript{-1}] \\
  $\sigma_a$ & Absorption coefficient & [m\textsuperscript{-1}] \\
  $\sigma_t = \sigma_s + \sigma_a$ & Extinction coefficient & [m\textsuperscript{-1}] \\
  $\rho = \sigma_s / \sigma_t$ & Scattering albedo & - \\
  \hline  
\end{tabular}
\caption{Parameters defining a sub-surface scattering material}
\label{tab:mat_params}
\end{table}

The goal of path tracing rendering is to estimate the rendering equation using Monte-Carlo integration.
\begin{equation*}
\begin{split} 
&L_o(p_o, \omega_o) = \\
&\int_A \int_{H^2(n)} S(p_o, \omega_o, p_i, \omega_i)L_i(p_i, \omega_i)\left|\cos\theta_i\right|\diff\omega_i\diff A
\end{split}
\end{equation*}
The challenge for sub-surface scattering materials is to be able to evaluate and importance sample the BSSRDF $S(p_o, \omega_o, p_i, \omega_i)$ given material parameters described in table~\ref{tab:mat_params}. In our approach, we make the simplifying assumption that the BSSRDF is factored into a spatial term $S_p$ and a directional term $S_\omega$. 
\begin{equation*} 
S(p_o, \omega_o, p_i, \omega_i) = (1-F_r(\cos\theta_o))S_p(p_o, p_i, \omega_i)S_\omega(\omega_i)
\end{equation*}
where $F_r$ is the Fresnel transmitance. Details about this decomposition can be found in PBRT 3\textsuperscript{rd} ed. section 15.4.1. $S_p$ can further be split into a single-scattering term $S_p^\mathrm{SS}$ and a multi-scattering  term $S_p^\mathrm{MS}$, accounting respectively for light paths with one and multiple internal bounces.
\begin{equation*} 
S_p = S_p^\mathrm{SS} + S_p^\mathrm{MS}
\end{equation*}
Our work focuses on building a model for the multi-scattering spatial term $S_p^\mathrm{MS}$.

\subsection{Photon Beam Diffusion}

\begin{figure}
\centering
\begin{subfigure}{.45\linewidth}
 	\centering
	\includegraphics[width=.63\textwidth]{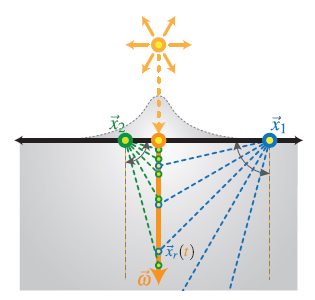}
 	\caption{Perpendicular incidence}
 	\label{fig:pbd_perp}
\end{subfigure}
\begin{subfigure}{.45\linewidth}
  	\centering
	\includegraphics[width=\textwidth]{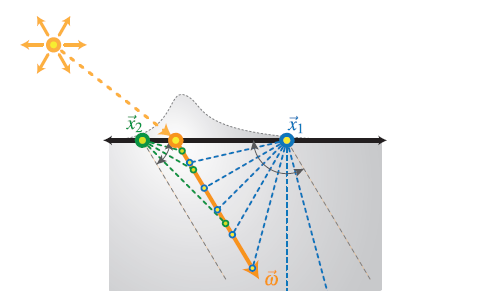}
  	\caption{Non-perpendicular incidence}
\end{subfigure}
\caption{Schematic representation of Photon Beam Diffusion. The multi-scattering BSSRDF at points $\vec{x_1}$ and $\vec{x_2}$ is computed as the sum of contributions of dipoles located along an internal ray. Images from of \protect\cite{habel2013photon}.}
\label{fig:pbd}
\end{figure}

We use Photon Beam Diffusion (PBD) as a reference to evaluate $S_p^\mathrm{MS}$. PBD is an extension of Jensen's dipole~\cite{jensen2001} that provides an approximation of the multi-scattering BSSRDF in a planar setting. Moreover, it naturally accounts for oblique incidence. In Jensen's original approach, the BSSRDF at some point of the surface is computed as the sum of contributions from a real light source (inside the medium) and a virtual light source (outside the medium). As shown in figure~\ref{fig:pbd}, PBD integrate the effects of an infinite number such dipoles along an internal ray, resulting in a analytic expression for $S_p^\mathrm{MS}$.

\begin{equation*}
S_{p}^\mathrm{MS}(\eta, g, \sigma_s, \sigma_a, \theta, r, \phi) = \int^\infty_0 (R^d_\phi(t) + R^d_{\vec{E}}(t))\kappa(t)Q(t)\diff t
\end{equation*}
where $\eta, g, \sigma_s, \sigma_a$ are the material parameters, $\theta = \theta_i$ and $(r, \phi)$ are the polar coordinates of $p_o$ relative to $p_i$. $\phi$ is measured with respect to the incidence plane. Table~\ref{tab:mat_params} describes the notations we use for material parameters and Figure~\ref{coor:model} illustrates the coordinate system. Details of the integrand can be found in appendix A. We use the technique described in PBRT-v3 section 15.5.6 to compute $S_p^\mathrm{MS}$ using numerical integration. 

\begin{figure}
%
%
%
%
%
\includegraphics{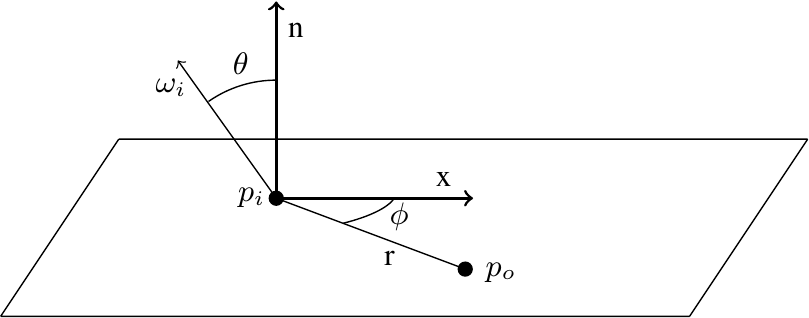}
\caption{The coordinate system we use in our model. Note that $\omega_i$, $n$ and $x$ are coplanar.}
\label{coor:model}
\end{figure}
 
\subsection{The perpendicular incidence assumption}

In order to improve convergence in a Monte-Carlo path tracer, one must be able to importance sample the BSSRDF. However, $S_p^\mathrm{MS}$ cannot be sampled directly because of its integral form. In practice, PBRT generates a tabulated version of the BSSRDF to be able to importance sample efficiently. 

As described in PBRT section 11.4.2, scaling the problem allows to assume $\sigma_t = 1$ , effectively reducing by one the dimensionality of $S_p^\mathrm{MS}$. Moreover, $\eta$ and $g$ do not need to be tabulated since they are usually constant for a given material. But tabulating the 4 remaining dimensions $(\rho, \theta, r, \phi)$ would produce a model too large for practical purposes. Thus PBRT makes the simplifying assumption that the incident light ray is perpendicular to the surface, as in figure~\ref{fig:pbd_perp}. This fixes $\theta = 0$ and removes $\phi$ variations, leaving only 2 dimensions to tabulate. PBRT then uses Catmull-Rom interpolation to evaluate and sample this tabulated model.

In our model, we account for all 4 dimensions $(\rho, \theta, r, \phi)$ by tabulating 3 of them $(\rho, \theta, r)$ and using an analytic approximation for the $\phi$ dimension.

\section{Building our non-perpendicular BSSRDF model}

\begin{figure}
\centering
\begin{subfigure}{.4\linewidth}
 	\centering
	\includegraphics[height=.5\textwidth]{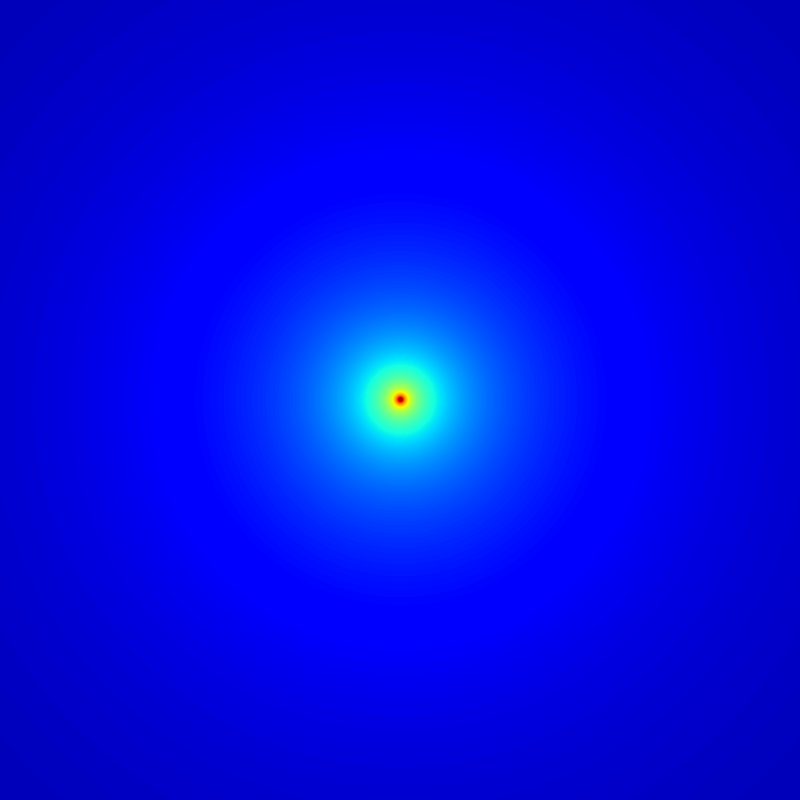}
 	\caption{Perpendicular incidence}
\end{subfigure}
\begin{subfigure}{.4\linewidth}
  	\centering
	\includegraphics[height=.5\textwidth]{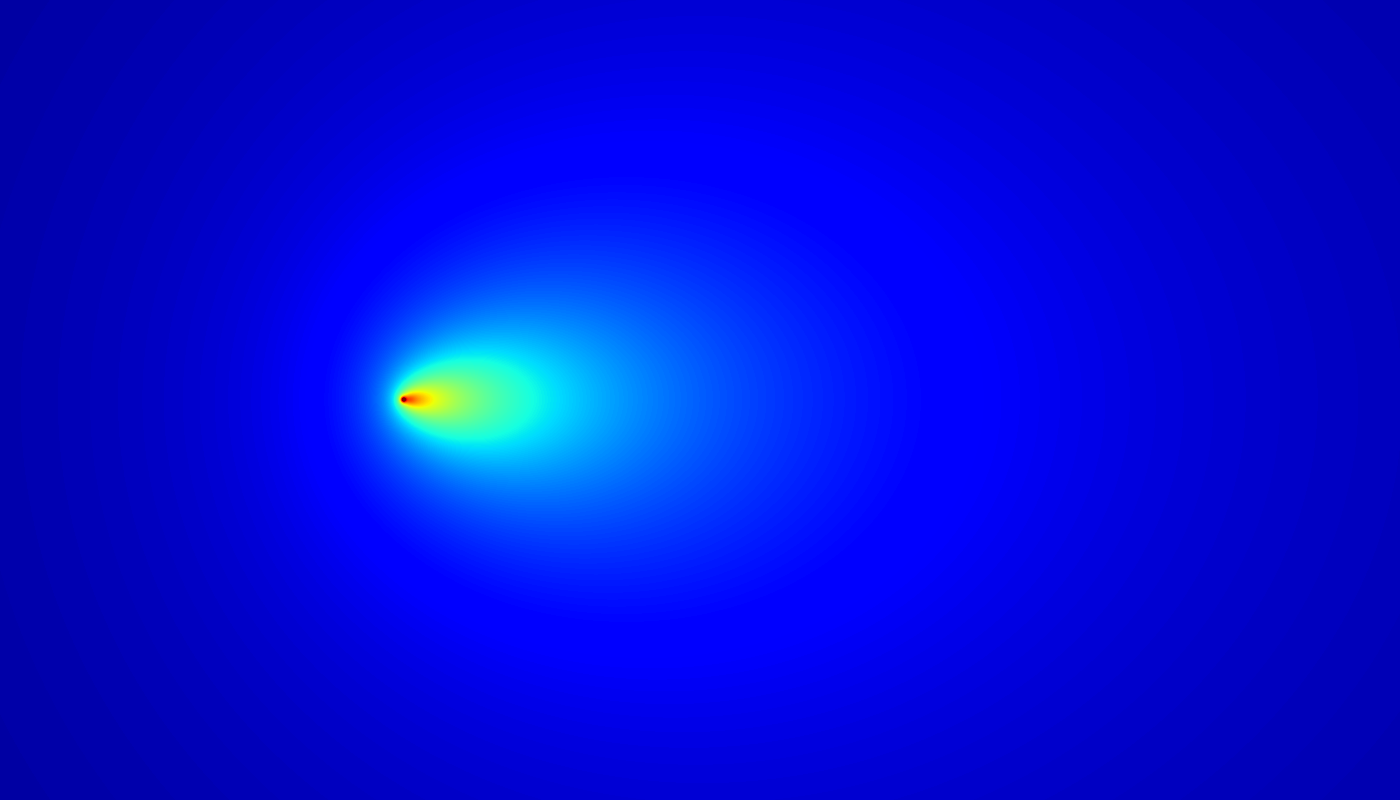}
  	\caption{89\degree incidence}
\end{subfigure}
\begin{subfigure}{.1\linewidth}
  	\centering
	\includegraphics[height=2.8\textwidth]{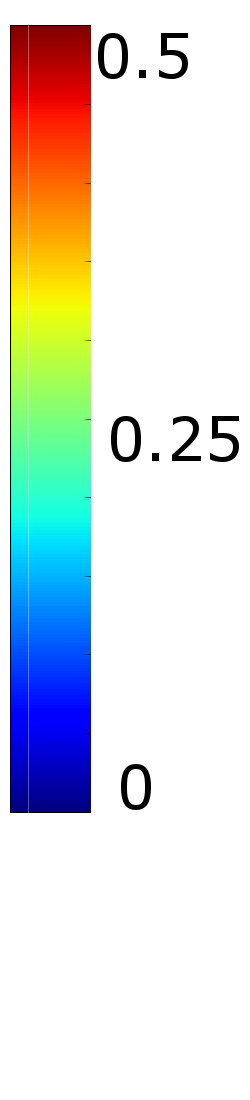}
\end{subfigure}

\caption{Comparison of the Photon Beam Diffusion BSSRDF for a light ray intersecting a semi-infinite slab of sub-surface scattering material at angles $\theta = 0\degree$ and $\theta = 89\degree$. Note the anistropic diffusion of light for non-perpendicular incidence. Material parameters: $\eta = 1.33$, $g = 0$, $\rho = 0.95$. }
\label{fig:incidence_heatmap}
\end{figure}

Anisotropic scattering effects produced by oblique rays are ignored under the perpendicular incidence assumption, as shown in Figure~\ref{fig:incidence_heatmap}.
In order to render those effects, our model does not make this assumption i.e. it accounts for the angle of incidence $\theta$ and the resulting non-uniform angular distribution of light ($\phi$ dimension). More precisely, it stores the function 

\begin{align*}
  	T\colon &[0, 1] \times [0,\frac{\pi}{2}] \times [0, +\infty] \times [-\pi, \pi] \to \mathbb{R}_+\\
  	&\rho, \theta, r, \phi \mapsto S_p^\mathrm{MS}(\eta, g, \rho, 1-\rho, \theta, r, \phi)
\end{align*}

for fixed $\eta > 1$ and $g \in [-1, 1]$. During path tracing, $\rho$ and $\theta$ are evaluated from the surface intersection. Then $r, \phi$ are jointly sampled according to the distribution defined by $T(\rho, \theta)$. In the following, we assume $\eta$ and $g$ fixed and explain the structure of our model. Then we explain how to build it using Photon Beam Diffusion evaluations.

\subsection{3D tabulation}

First, we use a 3D table of dimension $(\rho,\theta, r)$ to tabulated the radial energy
\begin{equation*}
E(\rho, \theta, r) = r\int_{-\pi}^{\pi} S^\mathrm{MS}_p(\rho, 1-\rho, \theta, r, \phi)\diff\phi
\end{equation*}
as well as parameters $\beta(\rho, \theta, r)$ and $c(\rho, \theta, r)$ of our angular model (described in the next subsection).
This 3D table is defined by the discretization we choose for $\rho,\theta$ and $r$ dimensions. We use the same discretization strategies as PBRT for $\rho$ and $r$.
\begin{equation*}
\rho_i = \frac{1 - e^{-8i / 99}}{1 - e^{-8}},\hspace{.2cm} i \in \{0, ..., 99\}\\
\end{equation*}
\begin{equation*}
r_0 = 0,\hspace{.5cm} r_i = 0.0025\times(1.2)^{i},\hspace{.2cm} i \in \{1, ..., 63\}
\end{equation*}
For $\theta$, we choose to use 10 uniform discretization points.
\begin{align*}
\theta_i = \frac{i\pi}{18},\hspace{.2cm} i \in \{0, ..., 9\}
\end{align*}

\subsection{Angular model}

\begin{figure*}
\centering
\begin{subfigure}{.49\linewidth}
 	\centering
	\includegraphics[height=.8\textwidth]{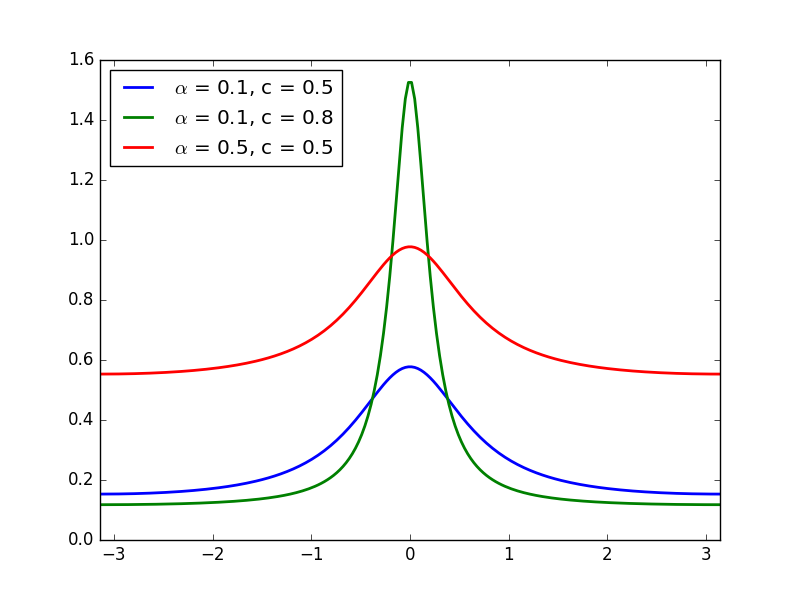}
 	\caption{}
 	\label{fig:gwc_eval}
\end{subfigure}
\begin{subfigure}{.49\linewidth}
  	\centering
	\includegraphics[height=.8\textwidth]{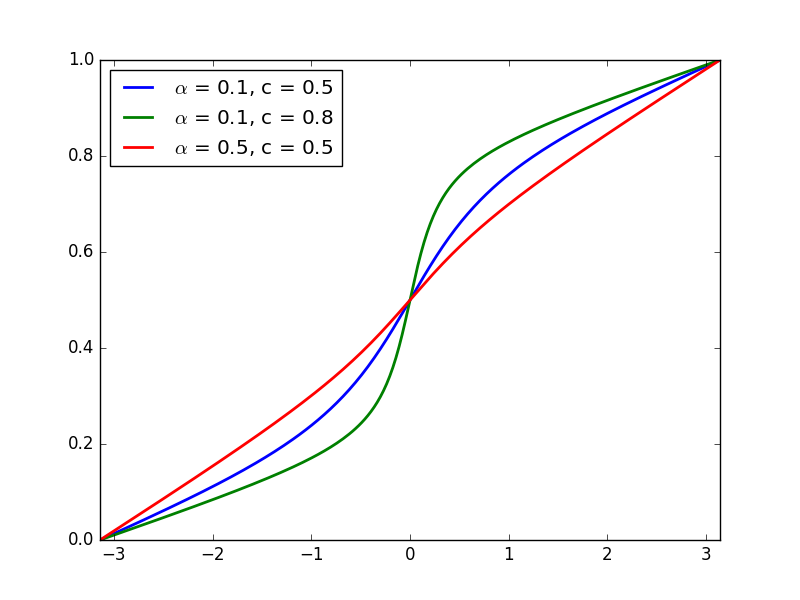}
  	\caption{}
 	\label{fig:gwc_cdf}
\end{subfigure}
\caption{General Wrapped Cauchy function (a) and corresponding cdf after normalization (b) for different $\alpha, c$ parameters and fixed $\beta = 1$.}
\label{fig:gwc}
\end{figure*}

For given $(\rho, \theta, r)$, we approximate $\phi \to S_p^\mathrm{MS}(\rho, 1-\rho, \theta, r, \phi)$ with a General Wrapped Cauchy (GWC) function $f_\mathrm{GWC}$ with 3 parameters $\alpha \geq 0, \beta \geq 0$ and $0 \leq c < 1$.
\begin{align*}
  	f_\mathrm{GWC}  \colon [-\pi, \pi] &\to \mathbb{R}_+\\
  	\phi; \alpha, \beta, c &\mapsto \alpha + \beta\,pdf_\mathrm{WC}(\phi; c)
\end{align*}
where WC is the Wrapped Cauchy distribution
\begin{align*}
  	pdf_\mathrm{WC} \colon [-\pi, \pi] &\to \mathbb{R}_+\\
  	\phi; c &\mapsto \frac{1}{2\pi}\frac{1-c^2}{1+c^2-2c\cos(\phi)}
\end{align*}

\begin{figure*}
\centering
\includegraphics[width=\textwidth]{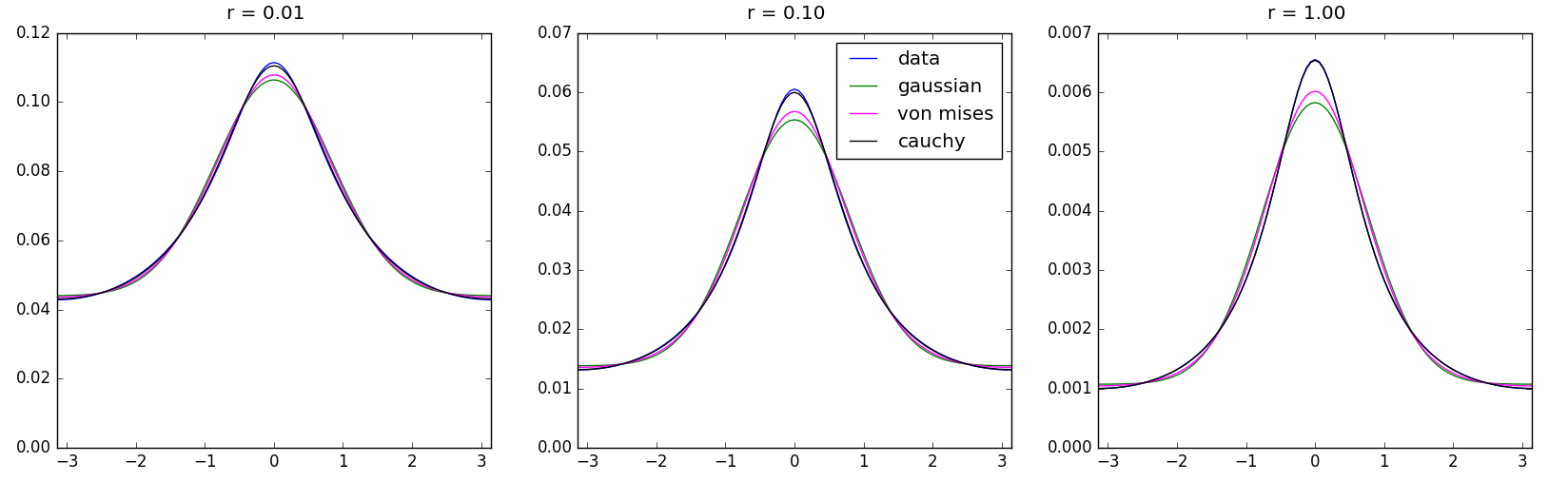}
\caption{Comparison of different angular models to fit angular variations of the multi-scattering BSSRDF. Photon Beam Diffusion data is plotted in blue. The models compared are : Wrapped Gaussian with an offset in green, Von Mises with an offset in pink, General Wrapped Cauchy in black. Model fitting was performed numerically using Scipy~\protect\cite{scipy}. The General Wrapped Cauchy model is consistently the best fit. Material parameters: $\eta = 1.33$, $g = 0$, $\rho = 0.5$, $\theta = 89\degree$. }
\label{fig:angular_fit}
\end{figure*}

Plots of $f_\mathrm{GWC}$ for different parameters are shown in figure~\ref{fig:gwc_eval}. We compared several angular models before choosing $f_\mathrm{GWC}$, as shown in figure~\ref{fig:angular_fit}. Note that $(E, \beta, c)$ can be computed from $(\alpha, \beta, c)$, and vice versa, using the equation
\begin{align*}
	E &= r\int_{-\pi}^{\pi} f_\mathrm{GWC}(\phi; \alpha, \beta, c)\diff\phi\\
 	&= (2\pi\alpha + \beta)\,r \numberthis \label{eq:energy}
\end{align*}

Thus storing $(E, \beta, c)$ values for every $(\rho, \theta, r)$ point of our table entirely describes our model. 

\subsection{Building our model}

\begin{figure*}
\centering
\begin{subfigure}{.49\linewidth}
 	\centering
	\includegraphics[width=\textwidth]{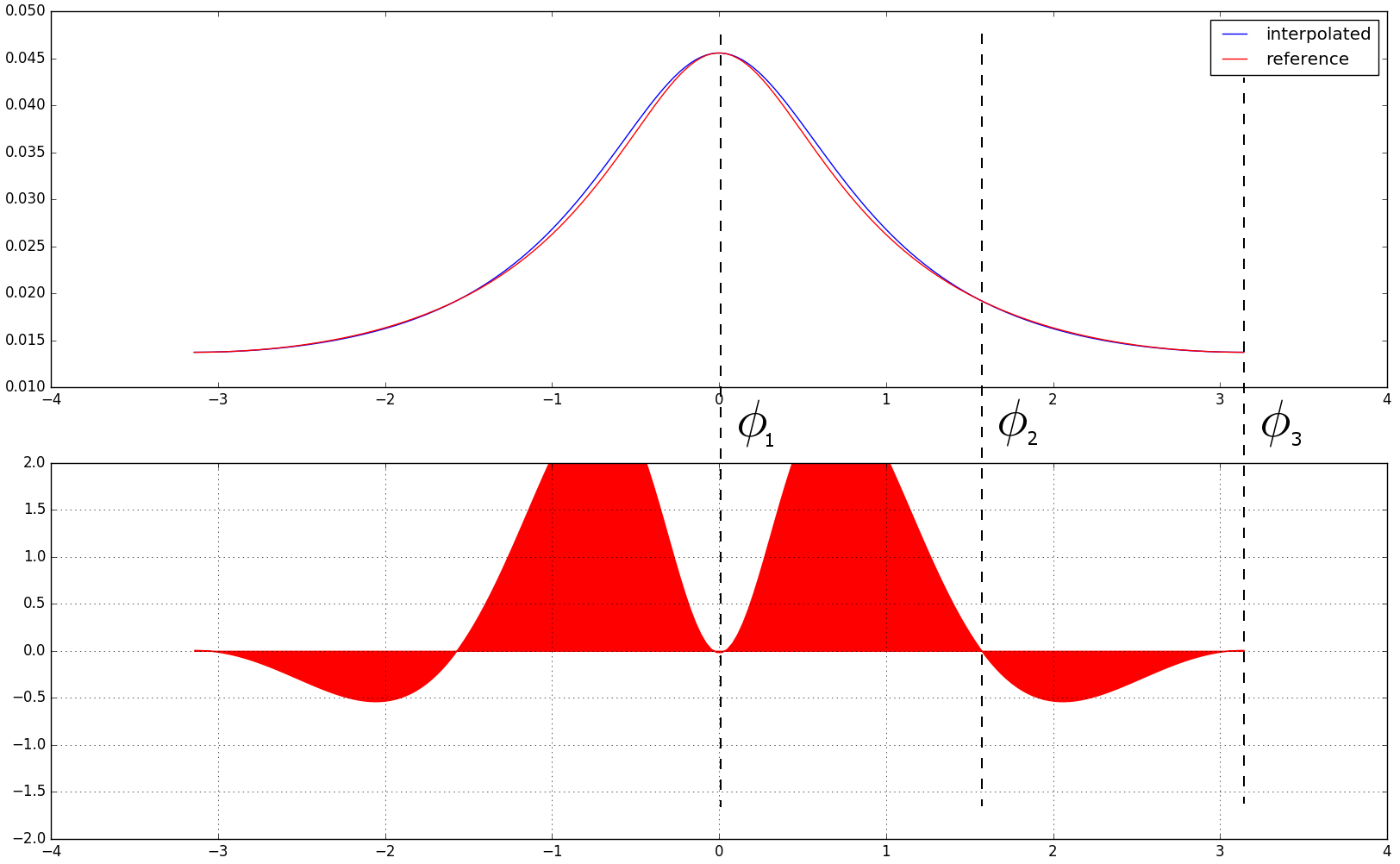}
  	\caption{Using $\cos\phi_1 = 1$, $\cos\phi_2 = 0$, $\cos\phi_3 = -1$}
\end{subfigure}
\begin{subfigure}{.49\linewidth}
  	\centering
	\includegraphics[width=\textwidth]{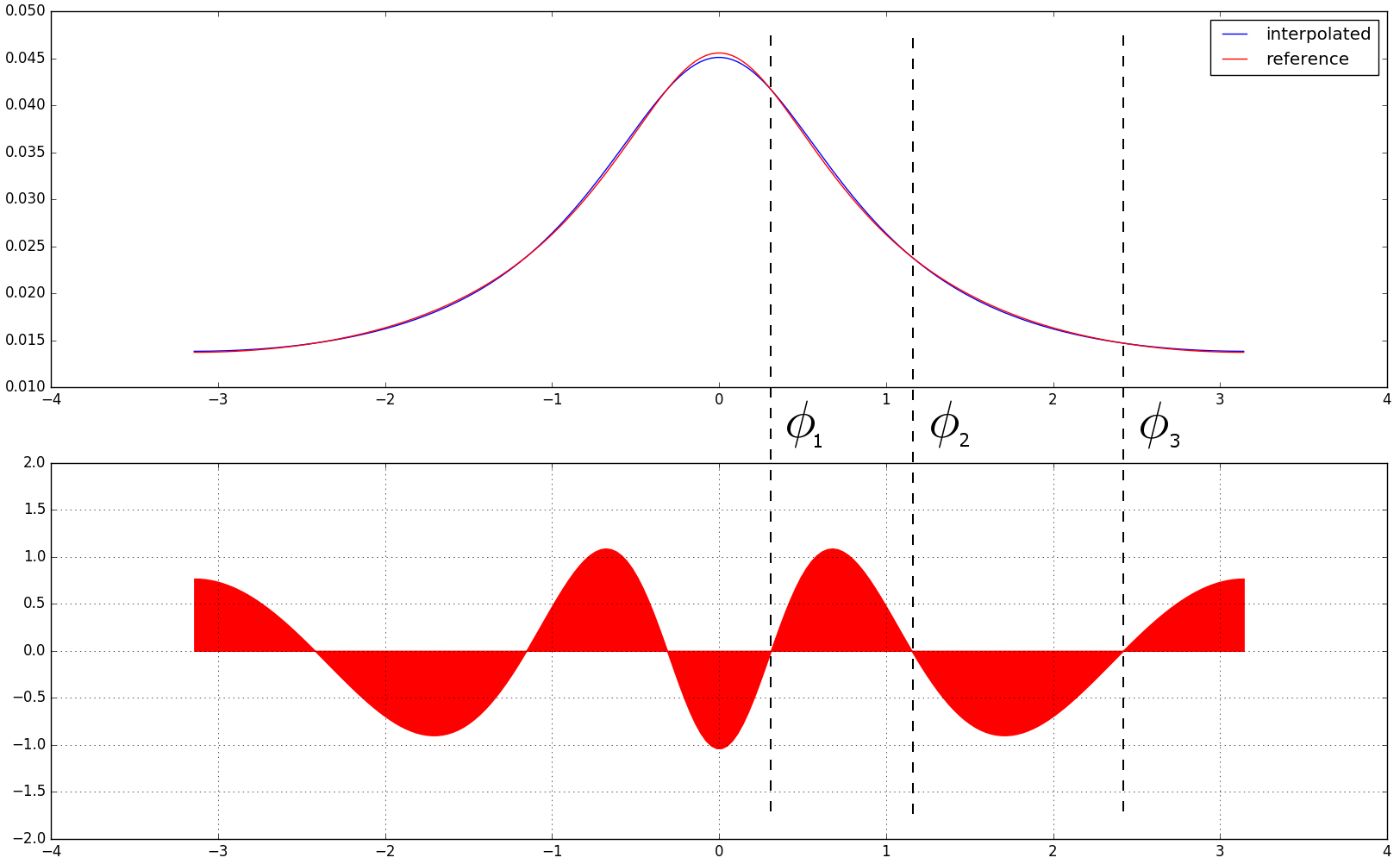}
 	\caption{Using $\cos\phi_1 = 0.9530$, $\cos\phi_2 = 0.4050$, $\cos\phi_3 = -0.7527$}
\end{subfigure}
\caption{Comparison of 2 sets of anchor points $(\phi_1, \phi_2, \phi_3)$ to fit our General Wrapped Cauchy model (blue plot, top) to Photon Beam Diffusion reference (red plot, top). Relative error (in $\%$) between our model and the reference is shown in the bottom row . Optimizing the location of the anchor points (right) allows to reduce the relative error of our model. Experiment parameters: $\eta = 1.33$, $g = 0$, $\rho = 0.99$, $\theta = 89\degree$, $r = 1$.}
\label{fig:anchors}
\end{figure*}

As in PBRT, our model is generated during scene preprocessing using Photon Beam Diffusion evaluations. For each $(\rho, \theta, r)$ sample in our table, we use 3 angular Beam Diffusion samples 
\begin{equation*}
f_i \underset{def}{=} S_p^\mathrm{MS}(\rho, 1-\rho, \theta, r, \phi_i),\hspace{.2cm} i \in \{1, 2, 3\}
\end{equation*}
to compute parameters $(E, \beta, c)$ that will model the angular profile. 
The anchor points $\phi_1, \phi_2, \phi_3$ impact the accuracy of our model as shown in figure~\ref{fig:anchors}. In practice we choose
\begin{align*}
	\cos\phi_1 &= 0.9530\\
	\cos\phi_2 &= 0.4050\\
	\cos\phi_3 &= -0.7527
\end{align*}

We first solve 
\begin{equation*}
\forall i \in \{1, 2, 3\},\hspace{.2cm} f_i = f_\mathrm{GWC}(\phi_i; \alpha, \beta, c) 
\end{equation*}
for $(\alpha, \beta, c)$ and then use equation \ref{eq:energy} to retreive $E$.

If $f_2 = f_3$, there exists no unique solution. We choose
\begin{empheq}[left=\empheqlbrace]{align*}
  	&c = 0\\
  	&\beta = 2\pi f_1\\
  	&\alpha = 0
\end{empheq}

Else, 
\begin{empheq}[left=\empheqlbrace]{align*}
  	&c = a - b\\
  	&\beta = \frac{2\pi(f_1-f_2)}{b}\bigg[\frac{1}{a - \cos\phi_1} - \frac{1}{a - \cos\phi_3}\bigg]^{-1}\\
  	&\alpha = f_1 - \frac{\beta\,b}{2\pi(a - \cos\phi_1)}
\end{empheq}
where
\begin{empheq}{align*}
  	&k = \frac{\cos\phi_1 - \cos\phi_2}{\cos\phi_2 - \cos\phi_3}\\
  	&K = \frac{f_1 - f_2}{f_2 - f_3}\\
  	&a = \frac{K\cos\phi_1 - k\cos\phi_3}{K-k}\\
  	&b = \sqrt{a^2-1}
\end{empheq}

\section{Manipulating our model}
There are two main operations that need to be efficiently achieved with our model: evaluation and importance sampling. In this section, we explain how to perform those two operations and measure the performance of our model in terms of memory footprint and evaluation accuracy. We rely partly on the Catmull-Rom interpolation and sampling algorithms described in PBRT.

\subsection{Evaluation}

Evaluation of our model for given $(\rho, \theta, r, \phi)$ is done in 2 steps. First, we use Catmull-Rom interpolation of the 3D table to compute the corresponding $(E, \beta, c)$ parameters. Then, we evaluate the GWC function defined by these parameters at angle $\phi$.

\begin{algorithm}
\caption{Evaluation}
\label{alg:evaluation}
\begin{algorithmic}[0] 
\Procedure{Evaluate}{$\rho, \theta, r, \phi$}
\State Get $(E, \beta, c)$ from the 3D table using interpolation
\State Set $\alpha = \frac{E/r - \beta}{2\pi}$
\State \textbf{return} $f_\mathrm{GWC}(\phi; \alpha, \beta, c)$
\EndProcedure
\end{algorithmic}
\end{algorithm}

\subsection{Importance sampling}

Importance sampling of $(r, \phi)$ for given $(\rho, \theta)$ is done in 3 steps. First, $r$ is sampled according to the radial energy distribution using Catmull-Rom sampling of the 3D table. Then parameters $(E, \beta, c)$ are evaluated at point $(\rho, \theta, r)$ of the 3D table using Catmull-Rom interpolation. Finally, $\theta$ is sampled according to the distribution defined by $f_\mathrm{GWC}$ with parameters $(\frac{E/r - \beta}{2\pi}, \beta, c)$.

For the last step, we rely the fact that $f_\mathrm{GWC}$ has an analytic cdf to perform Newton-Bisection.
\begin{align*}
 	cdf_\mathrm{GWC} \colon [-\pi, \pi] &\to [0, 1]\\
  	\phi; \alpha, \beta, c &\mapsto \frac{\alpha(\phi+\pi) + \beta\,cdf_\mathrm{WC}(\phi; c)}{2\pi\alpha + \beta}
\end{align*}
where
\begin{align*}
  	cdf_\mathrm{WC} \colon [-\pi, \pi] &\to [0, 1]\\
 	\phi; c &\mapsto \frac{1}{2} + \frac{1}{\pi}\arctan\Big(\frac{1+c}{1-c}\tan(\frac{\phi}{2})\Big)
\end{align*}
Plots of $cdf_\mathrm{GWC}$ for different parameters are shown in figure~\ref{fig:gwc_cdf}. Since in practice $2\pi\alpha$ is small compared to $\beta$, we initiate Newton-Bisection using the analytic Wrapped Cauchy inverse cdf (i.e. assuming $\alpha = 0$).
\begin{align*}
  	cdf^{-1}_\mathrm{WC} \colon [0, 1] &\to [-\pi, \pi]\\
 	x; c &\mapsto 2\arctan\Big(\frac{1-c}{1+c}\tan(\pi(x-\frac{1}{2})\Big)
\end{align*}

\begin{algorithm}
\caption{Importance sampling} \label{alg:sampling}
\begin{algorithmic}[0]
\Procedure{Sample}{$\rho, \theta$}
\State Sample $r$ using Catmull-Rom sampling of the 3D table
\State Get $(E, \beta, c)$ from the 3D table using interpolation
\State Sample $\phi$ using Newton-Bisection of $cdf_\mathrm{GWC}$
\State \textbf{return} $(r, \phi)$
\EndProcedure
\end{algorithmic}
\end{algorithm}

\subsection{Performance}

Since our model is partly based on interpolation, there is a trade-off between memory footprint and accuracy. Memory-wise our model is a 3D table of size $100 \times 10 \times 64$, respectively for the $\rho, \theta, r$ dimensions. Each cell stores 3 parameters: $(E, \beta, c)$. Additionally, we use a 3D table to store the cumulative radial energy $\int_0^r E(r')\diff r'$ in order to optimize Catmull-Rom sampling. Thus our BSSRDF model takes 1 MiB of storage if numbers are encoded on 4 bytes using single precision floating point values. 

\begin{figure*}
\centering
\begin{subfigure}{.3\linewidth}
 	\centering
	\includegraphics[width=\textwidth]{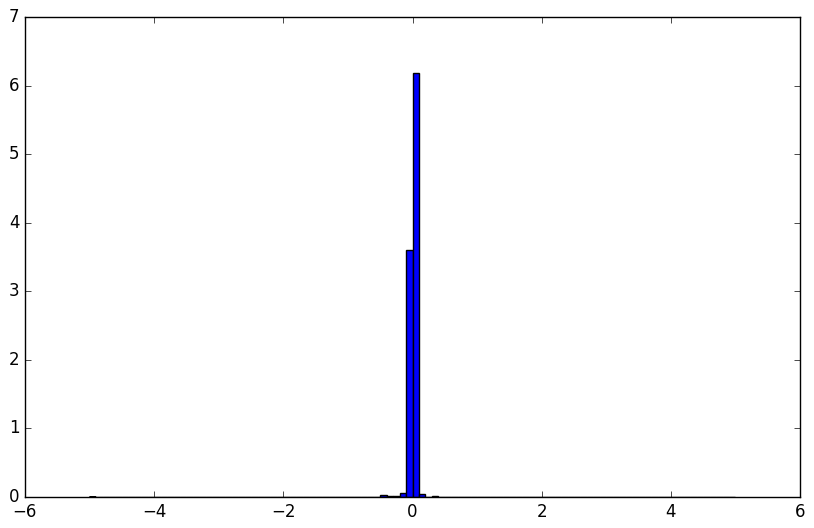}
\end{subfigure}
\begin{subfigure}{.3\linewidth}
  	\centering
	\includegraphics[width=\textwidth]{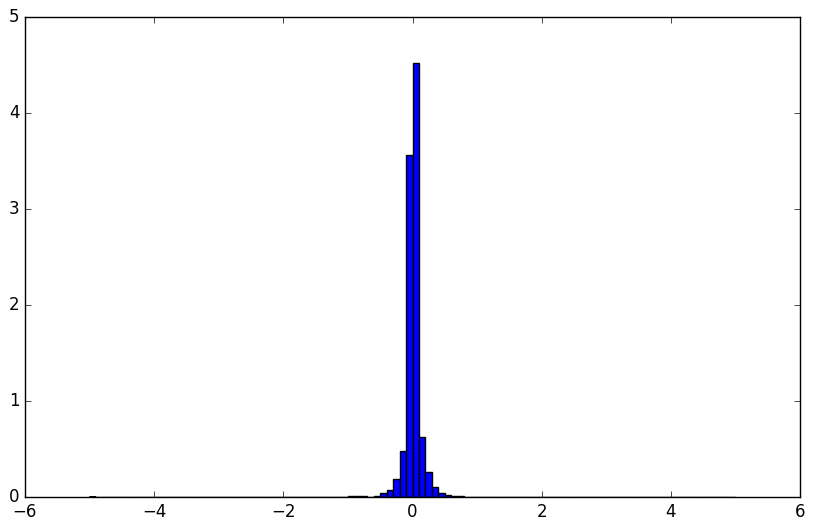}
\end{subfigure}
\begin{subfigure}{.3\linewidth}
  	\centering
	\includegraphics[width=\textwidth]{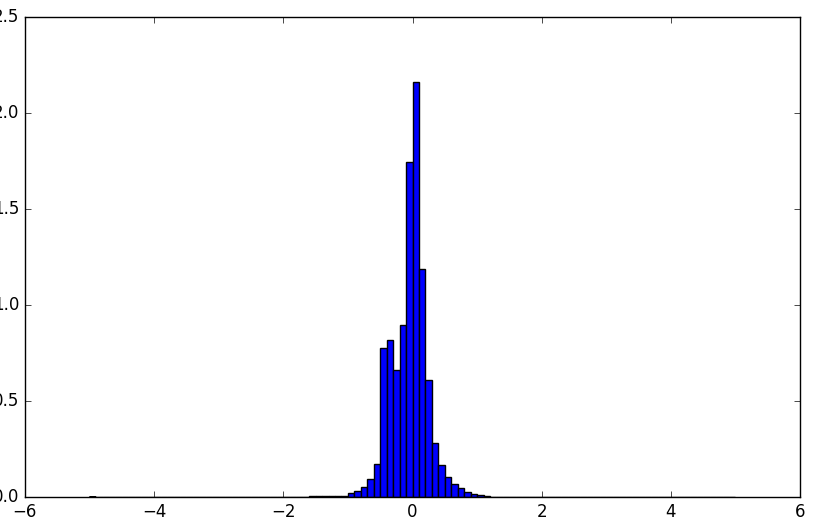}
\end{subfigure}
\begin{subfigure}{.3\linewidth}
 	\centering
	\includegraphics[width=\textwidth]{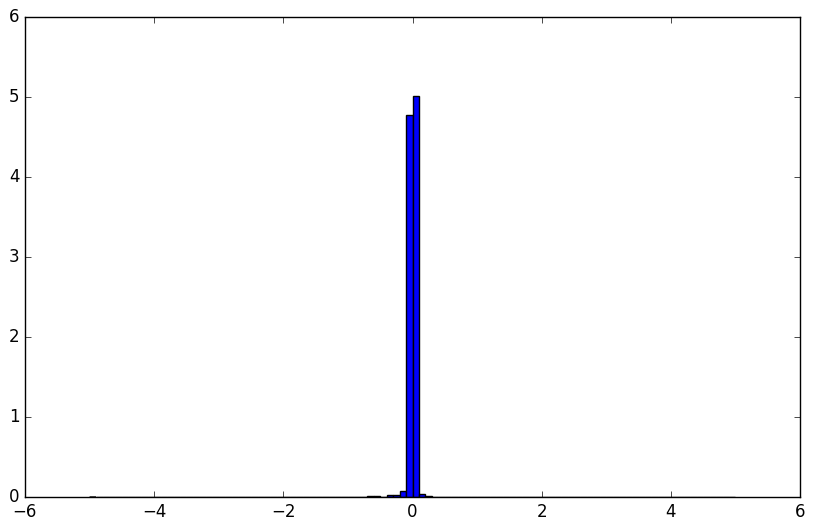}
  	\caption{$\theta = 0\degree$}
\end{subfigure}
\begin{subfigure}{.3\linewidth}
  	\centering
	\includegraphics[width=\textwidth]{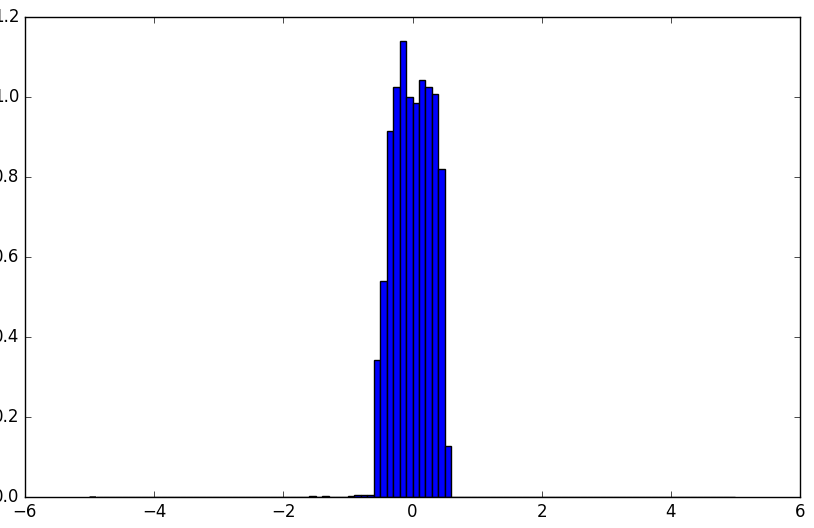}
 	\caption{$\theta = 60\degree$}
\end{subfigure}
\begin{subfigure}{.3\linewidth}
  	\centering
	\includegraphics[width=\textwidth]{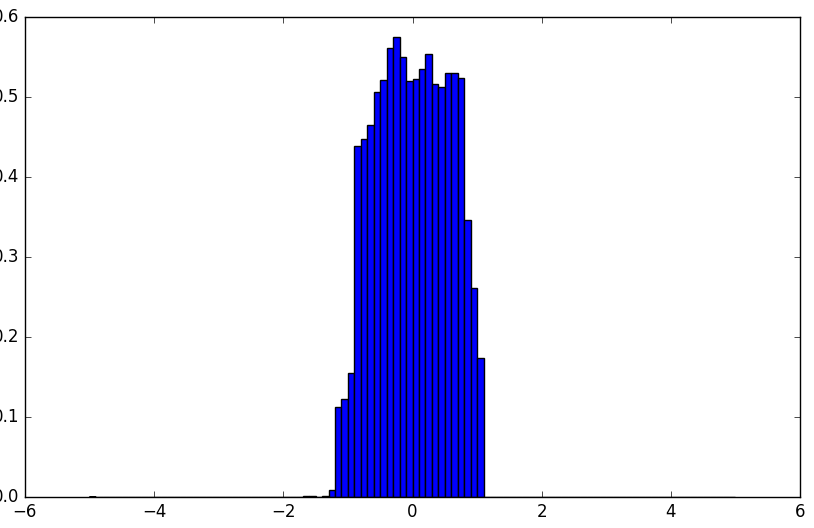}
 	\caption{$\theta = 89\degree$}
\end{subfigure}

\caption{Histograms of relative error ($\%$) between evaluation using our model and Photon Beam Diffusion reference. Material parameters for these measures are $\eta = 1.33$, $g = 0$, $\rho = 0.5$ (top) and $\rho = 0.99$ (bottom). To generate those histograms, 100000 points were sampled using algorithm~\ref{alg:sampling}. These results shows that our model produces below $1\%$ uniform relative error, even for grazing angles and high albedo.}
\label{fig:rel_error}
\end{figure*}

Evaluation of $S_p^\mathrm{MS}$ using our model is within $0.5\%$ mean relative error from the Photon Beam Diffusion reference. This holds even for grazing angles, as shown in table~\ref{tab:mean_error}. Moreover the relative error distribution produced by importance sampling is within $1\%$ uniform relative error of the reference, as shown in figure~\ref{fig:rel_error}.

\begin{table}
\centering
\begin{tabular}{ l c c c}
  \toprule
    & \multicolumn{3}{c}{Mean relative error ($\%$)}\\
    \cmidrule{2-4}
    & $\rho = 0.5$ & $\rho = 0.9$ & $\rho = 0.99$ \\
  \midrule
   $\theta = 0\degree$  & $0.026$ & $0.026$ & $0.021$\\
   $\theta = 60\degree$ & $0.08$ & $0.26$ & $0.25$\\
   $\theta = 89\degree$ & $0.22$ & $0.53$ & $0.48$\\
  \bottomrule
\end{tabular}
\caption{Mean relative error of our model for different $(\rho, \theta)$ values. 100000 samples were generated using importance sampling (algorithm~\ref{alg:sampling}) to compute each mean.}
\label{tab:mean_error}
\end{table}

\section{Implementation in PBRT}

By using a model of multiple scattering for oblique incidence, we are able to render anisotropic effects. In this section, we explain how to use our model in a particle tracer. We then show images rendered using our model and compare them to images rendered with perpendicular incidence.

\subsection{Geometry sampling}

While our method provides an accurate model of multi-scattering for a planar surface, it must be adapted in order to sample points of the scene geometry. We use the same ray casting technique as described in PBRT section 15.4.1. That is, we randomly choose the ray direction among 3 possible axis and then use the polar coordinates $(r, \phi)$ sampled with our model to set the position of the ray. However, if the axis chosen is not the one normal to the surface intersection, it makes little sense to use the $\phi$ sampled with our model. Thus we replace $\phi$ by $\phi'$ uniformly sampled in $[-\pi, \pi]$ in this case.

\subsection{Particle tracing results}

\begin{figure}
\centering
 	\begin{subfigure}[b]{0.22\textwidth}
        \includegraphics[width=.95\textwidth]{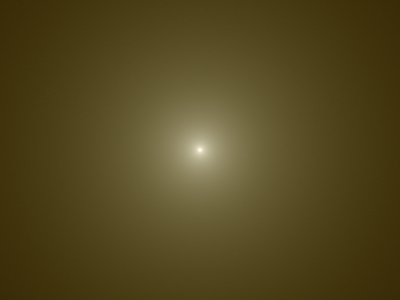}
        \caption{$\theta = 0\degree$}
    \end{subfigure}
    \begin{subfigure}[b]{0.22\textwidth}
        \includegraphics[width=.95\textwidth]{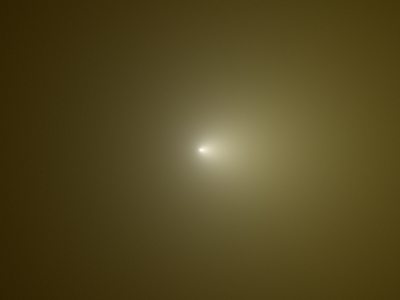}
        \caption{$\theta = 60\degree$}
    \end{subfigure}

\caption{Renderings of a concentrated cone of light intersecting a semi-infinite slab of sub-surface scattering material, using our model. These images show the anisotropic effect of non-perpendicular incidence of angle $\theta$. Rendered using a particle tracer in PBRT. Material parameters: $\eta = 1.33$, $g = 0$, $\sigma_s = 1$, $\sigma_a = (0.01, 0.1, 1)$.}
\label{fig:render_beam}
\end{figure}

Our method models light exiting the surface when an incident light ray is scattered. Thus it adapts naturally to particle tracing. We have implemented a simple particle tracer in PBRT and rendered scenes using our model.
Figure~\ref{fig:render_beam} shows renderings of a beam scattering on a infinite slab of sub-surface scattering material.
Our model produces anisotropic light scattering effects that cannot be rendered under the perpendicular incidence assumption, as illustrated in Figure~\ref{render_bunny}.

%

\section{Conclusion}

We present a compact model for multiscattering BSSRDFs. It uses Photon Beam Diffusion as reference and relies on  tabulation to store a model of angular variations. It allows fast evaluation and importance sampling with $0.5\%$ relative accuracy for a storage cost of 1 MiB. Using our model in a particle tracer allows to render anisotropic effects that would not appear under the perpendicular incidence assumption.

\section{Future work}

Particle tracing by itself is not a very efficient rendering technique. In practice our model would have to be integrated into a path tracer. In this section, we describe what remains to be done in order to use our model for path tracing. Moreover, we have explored using the same technique to build a model of single-scattering. We present the results of this study.

\subsection{Using our model for path tracing}

In order to use our model for path tracing, sampling needs to be reversed. That is, we need to importance sample $(p_i$, $\omega_i)$ from $(p_o$, $\omega_o)$. Note that algorithm~\ref{alg:evaluation} can be used for evaluation in both path tracing and particle tracing. 

Importance sampling of a new path can be done in 3 steps. First, $\theta_i$ is sampled from the effective albedo distribution 
\begin{align*}
\rho_\mathrm{eff} \colon [0, \frac{\pi}{2}] &\to \mathbb{R}_+ \\
\theta; \rho &\mapsto \int_0^\infty E(\rho, \theta, r)\diff r
\end{align*} Then a point $(r, \phi)$ is sampled using algorithm~\ref{alg:sampling} for $(\rho, \theta_i)$. This point defines $p_o$ in the polar coordinate system defined by $(p_i, \omega_i)$. Finally the azimuth $\phi'$ of $p_i$ relative to $p_o$ is sampled uniformly. Because the exiting light is assumed diffuse, $\theta_o$ does not appear in the process. 

\begin{figure}
%
%
%
%
\includegraphics{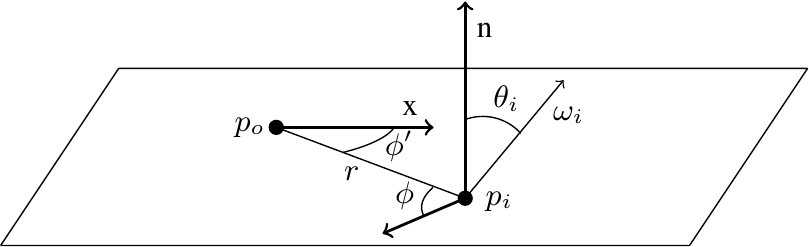}
\caption{Coordinate system used for importance sampling of a new path during path tracing.}
\label{coor:path}
\end{figure}

Figure~\ref{coor:path} illustrates the different geometric parameters in a planar setting and algorithm~\ref{alg:sampling_path} gives pseudo code for this procedure. 

\begin{algorithm} 
\caption{Importance sampling for path tracing}
\label{alg:sampling_path}
\begin{algorithmic}[0]
\Procedure{Sample}{$\rho$}
\State Sample $\theta_i$ using Catmull-Rom sampling of a 2D table.
\State Get $(r, \phi) =$ \Call{Sample}{$\rho, \theta_i$}
\State Sample $\phi'$ uniformly
\State \textbf{return} $(r, \phi', \theta_i, \phi)$
\EndProcedure
\end{algorithmic}
\end{algorithm}

\begin{figure}
%
%
%
%
\includegraphics{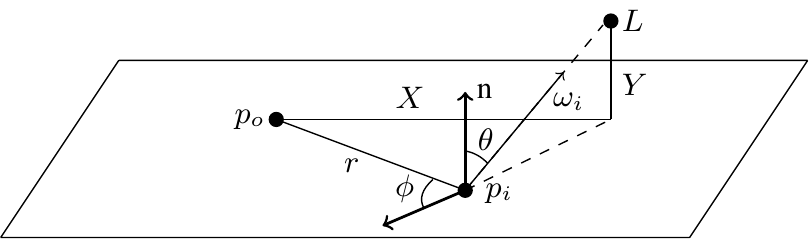}
\caption{Coordinate system used for importance sampling light sources during path tracing.}
\label{coor:lights}
\end{figure}

However, importance sampling light sources cannot be done directly with our model. Notably because it adds a non-linear constraint between the dimensions of our tabulated model.
\begin{equation*}
X^2 = r^2 + Y^2\tan^2\theta + 2rY\tan\theta\cos\phi
\end{equation*}
where $X$ is the distance from $p_o$ to the projection of the light source on the surface plane, and $Y$ is distance from the light source to the surface plane. Figure~\ref{coor:lights} illustrates these coordinates.

\subsection{Taking single scattering into account}
\label{sec:single-scatt}

\begin{figure*}
\centering
\includegraphics[width=\textwidth]{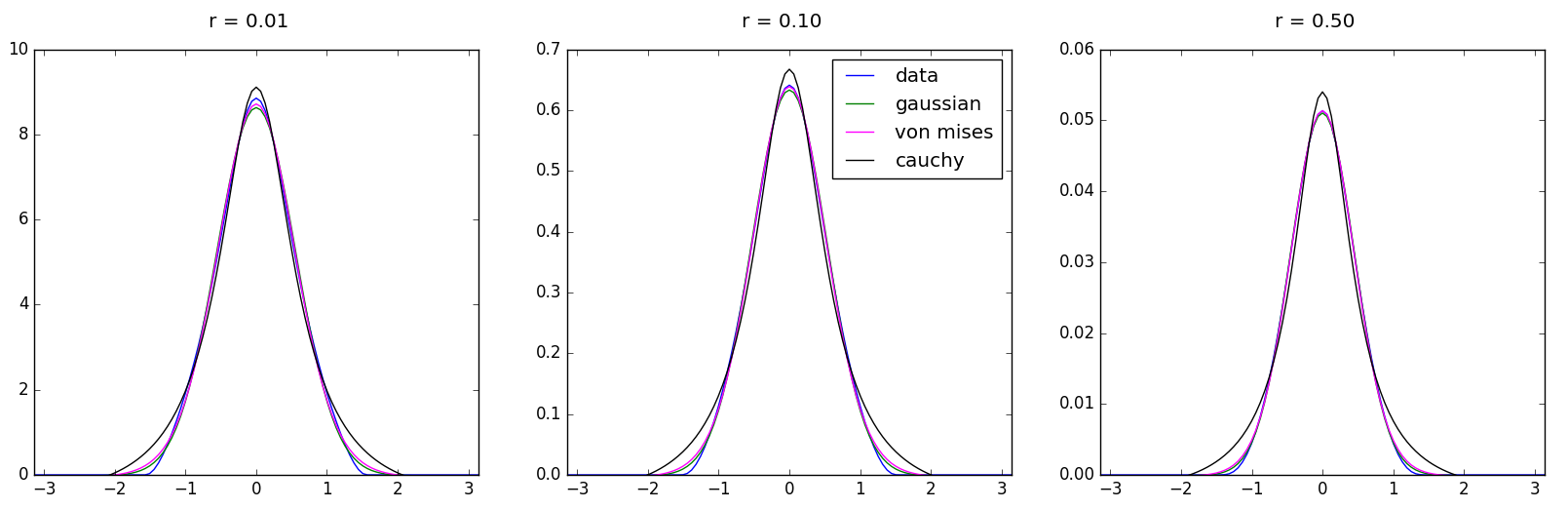}
\caption{Comparison of different angular models to fit angular variations of the single-scattering BSSRDF. Reference data computed using~\protect\cite{habel2013photon} is plotted in blue. The models compared are : Wrapped Gaussian  in green, Von Mises in pink, General Wrapped Cauchy in black. Model fitting was performed numerically using Scipy~\protect\cite{scipy}. Wrapped Gaussian and Von Mises models are the best fits. But despite being close to the reference in absolute error, they produce a high relative error because they cannot reproduce the sharp turn around $|\phi| = 1.5$. Material parameters: $\eta = 1.33$, $g = 0$, $\rho = 0.5$, $\theta = 89\degree$. }
\label{fig:angular_fit_ss}
\end{figure*}

Our model only accounts for multi-scattering. One way to also render single scattering would be to choose to sample multi or single-scattering on a coin flip. Single scattering can be sampled using volumetric path tracing for instance. However, it would be useful to jointly sample multi-scattering and single-scattering. That is the approach taken by PBRT but under the perpendicular incidence assumption. We have tried using our technique to model non-perpendicular single-scattering but with unsatisfying results as we show in Figure~\ref{fig:angular_fit_ss}. One issue we encountered is that the angular single-scattering profile is not differentiable at every angle because of total internal reflection, which makes it hard to fit with a smooth model.

\subsection{Additional remarks}

We some perspective, it would be clearer to change the parameters of the General Wrapped Cauchy function to
\begin{align*}
  	f_\mathrm{GWC}  \colon [-\pi, \pi] &\to \mathbb{R}_+\\
  	\phi; E, \lambda, c &\mapsto E\big[\frac{\lambda}{2\pi} + (1-\lambda)\,pdf_\mathrm{WC}(\phi; c)\big]
\end{align*}
instead of using two sets of notations: $(\alpha, \beta, c)$ and $(E, \beta, c)$.

\appendix
\section{Appendix: Photon Beam Diffusion}
\label{ape:pbd}
Photon Beam Diffusion provides an analytic expression of the multiscattering BSSRDF.

\begin{equation*}
S_{p}^\mathrm{MS}(\eta, g, \sigma_s, \sigma_a, \theta, r, \phi) = \int^\infty_0 (R^d_\phi(t) + R^d_{\vec{E}}(t))\kappa(t)Q(t)\,dt
\end{equation*}

where $\eta, g, \sigma_s, \sigma_a$ are the material parameters, $\theta$ is the incidence angle of the light ray before refraction and $(r, \phi)$ is the location of the exit point in polar coordinates. The azimuth $\phi$ is measured relative to the incidence plane.

Tables~\ref{tab:pbd_symbols} and \ref{tab:pbd_equ} detail the terms in the integrand for fixed $\eta, g, \sigma_s, \sigma_a, \theta, r, \phi$ values.

\begin{table}
\centering
\begin{tabular}{ c l c }
  \hline
  \vspace{-0.3cm}\\
  Symbol & Description & Units \\
  \hline
  $\Phi(\vec{x})$ & Fluence & [W m\textsuperscript{-2}] \\
  $\vec{E}(\vec{x})$ & Vector flux (vector irradiance) & [W m\textsuperscript{-2}] \\
  $Q(t)$ & Source function & [W m\textsuperscript{-3}] \\
  $\kappa(t)$ & Correction factor & - \\
  $F_1(\eta)$ & First Fresnel moment & - \\
  $F_2(\eta)$ & Second Fresnel moment & - \\
  \hline  
\end{tabular}
\caption{Symbols used in Photon Beam Diffusion equations}
\label{tab:pbd_symbols}
\end{table}

\begin{table}
\centering
\begin{tabular}{p{7cm}}
  \toprule
  \multicolumn{1}{c}{Material constants} \\
  \midrule
  \begin{equation*}
  \sigma_s' = \sigma_s(1-g)
  \end{equation*}
  \begin{equation*}
  \sigma_t' = \sigma_s' + \sigma_a
  \end{equation*}
  \begin{equation*}
  \rho' = \frac{\sigma_s'}{\sigma_t'}
  \end{equation*}
  \begin{equation*}
  \sigma_{tr} = \sqrt{\frac{\sigma_a}{D}}
  \end{equation*}
  \begin{equation*}
  D = \frac{2\sigma_a + \sigma'_s}{3\sigma'^2_t}
  \end{equation*}
  \begin{equation*}
  z_b = -2\,D\frac{1+3\,F_2(\eta)}{1-2\,F_1(\eta)}
  \end{equation*}
  \begin{equation*}
  C_\Phi = \frac{1-2\,F_1(\eta)}{4}
  \end{equation*}
  \begin{equation*}
  C_{\vec{E}} = \frac{1-3\,F_2(\eta)}{2}
  \end{equation*}\\
  \midrule
  \multicolumn{1}{c}{Geometric terms} \\
  \midrule
  \begin{equation*}
  \sin(\theta') = \sin(\theta) / \eta  \end{equation*}
  \begin{equation*}
  \cos(\theta') = \sqrt{1 - \sin^2(\theta')} 
  \end{equation*}
  \begin{equation*}
  z_r(t) = t\cos(\theta')
  \end{equation*}
  \begin{equation*}
  z_v(t) = 2\,z_b - z_r(t)
  \end{equation*}
  \begin{equation*}
  \lambda^2(t) = r^2 + t^2\sin^2(\theta') - 2\,r\,t\sin(\theta')\cos(\phi)
  \end{equation*}
  \begin{equation*}
  d_r(t) = \sqrt{\lambda^2(t) + z_r^2(t)}
  \end{equation*}
  \begin{equation*}
  d_v(t) = \sqrt{\lambda^2(t) + z_v^2(t)}
  \end{equation*}
  \begin{equation*}
  Q(t) = \rho' \sigma'_t e^{-\sigma'_t t}
  \end{equation*}\\
  \midrule
  \multicolumn{1}{c}{Fluence and vector irradiance contributions} \\
  \midrule
	\begin{equation*}
	R^d_\Phi(t) = C_\phi \frac{\rho'}{4\pi D} \bigg(\frac{e^{-\sigma_{tr}d_r(t)}}{d_r(t)} - 			\frac{e^{-\sigma_{tr}d_v(t)}}{d_v(t)}\bigg)
	\end{equation*}
	\begin{equation*}
	\begin{split}
	R^d_{\vec{E}}(t) =  C_{\vec{E}} \frac{\rho'}{4\pi} \bigg(\frac{z_r(t)(1+\sigma_{tr}d_r(t))e^{-	\sigma_{tr}d_r(t)}}{d_r^3(t)}+\\ \frac{(z_r(t)+2z_b)(1+\sigma_{tr}d_v(t))e^{-\sigma_{tr}d_v(t)}}	{d_v^3(t)}\bigg)
	\end{split}
	\end{equation*} \\
  \bottomrule
\end{tabular}
\caption{Detailed terms used in Photon Beam Diffusion}
\label{tab:pbd_equ}
\end{table}

\bibliographystyle{acmsiggraph}
\nocite{*}
\bibliography{template}
\end{document}